\documentclass[a4paper,twocolumn,preprintnumbers,amsmath,amssymb]{revtex4}
\usepackage{bm}
\usepackage{graphicx}

\begin{document}

\title{Robustness of  One-Dimensional Photonic Bandgaps Under Random Variations of Geometrical Parameters}

\author{H. Sami S\"{o}z\"{u}er and Koray Sevim}

\address{Izmir Institute of Technology, Department of Physics, Gulbahce Koyu, Urla, Izmir, TURKEY}

\email{sozuer@photon.iyte.edu.tr}

\begin{abstract}
The supercell method is used to study the variation of the
photonic bandgaps in one-dimensional photonic crystals
under random perturbations to thicknesses of the layers.
The results of both plane wave and analytical band structure and 
density of states calculations are presented along with the 
transmission coefficient as the level of randomness and the
supercell size is increased. 
It is found that higher bandgaps disappear first as the randomness is
gradually increased. The lowest bandgap is found to persist up to a 
randomness level of $55$ percent. 
\end{abstract}

\maketitle

\section{Introduction}

Since the pioneering work of E. Yablonovitch 
\cite{Yablonovitch} and S. John \cite{John}, research on photonic
crystals(PCs) has enjoyed a nearly exponential increase. The
manufacture of PCs at the optical regime has become
a reality \cite{lin}. Manufacturing brings with it the practical
reality of random errors introduced during the manufacturing
process and it is the effect of these random errors on the
desirable features of PCs, namely photonic band gaps, that we
wish to address in this paper.

Bandgaps in PCs depend on two crucial properties: an
infinite {\em and} perfect translational symmetry. Clearly, in real life no
crystal is infinite in size or perfectly periodic. 
When randomness is introduced in the geometry of the PC, one quantity
of interest is the size of the bandgaps as the level of randomness
is increased, and whether the bandgaps of the bulk perfect PC will survive
the randomness. Same considerations apply for a {\em finite} PC. In fact,
even for a perfect but finite PC, one needs to give up the notion of
a bandgap and has to be content with severe depressions in transmittance instead. 
In this paper, we will consider both finite imperfect PCs by examining the 
dependence of their transmittance on randomness,
and bulk imperfect PCs by determining their density of states (DoS) under 
varying degrees of randomness, using the supercell method.

Although much has been done \cite{lizhang2000,lizhangzhang2000,fanville95,sigalas,kaliteevski2000}
regarding imperfect two- and 
three-dimensional PCs, we feel that a study of the problem for  
one-dimensional PCs is warranted because of the inherent simplicity
of the geometry and because a variety of extremely accurate mathematical 
tools are readily available which allow a detailed study of the problem  without 
having to compromise accuracy. For instance, because the electric field
and its first derivative are continuous across the interface, and because of the
low dimensionality of the PC, the convergence problem 
that plagued band structure calculations for many 3D PCs \cite{sozuer,moroz} is 
essentially non-existent for 1D structures. Thus, we were able to use
the old trusted plane wave (PW) method to find the band
structure and the DoS for supercell sizes not even imaginable in 
three- or even two-dimensional supercell calculations \cite{lizhang2000,lizhangzhang2000}. 
One can obtain
better than 0.1\%  convergence with as few as $\sim$30 plane waves per unit cell
in the supercell. The transmission coefficient, too, can be calculated for
nearly arbitrary supercell sizes. Finally, one can calculate the band structure
and the imaginary part of the wave vector using a semi-analytical approach for 
very large supercells.

The 1D PC is, in many ways, the ``infinite square-well" problem of photonic crystals.
It contains the essential features of its bigger cousins in two and three dimensions
without the mathematical complexities and the accompanying numerical 
uncertainties \cite{sozuer,moroz} that can sometimes overshadow the
essentials. For example, with 3D face centered cubic structures, it becomes 
practically impossible, due to convergence problems, to
increase the supercell size beyond $2\times 2\times 2$ (or at most $3\times 3\times 3$) 
conventional cubic unit cells
which contain only 32 primitive cells per supercell (or 108), since typically at least 
$\sim 1000$ terms
per primitive cell are necessary to ensure sufficient convergence for inverse opal structures. 
It's not obvious from the start whether
a randomness analysis with such small supercell sizes would yield results that
are physically meaningful. Artifacts due to the small supercell size are bound 
to be inextricably intertwined with the  physically significant bulk features of the
imperfect PC. It's important to realize that, with the supercell method 
one still calculates the bands of
an infinite {\em perfect} PC. The randomness is only {\em within} the supercell,
but on a global scale, it's still a perfectly periodic structure! In order to 
be able to resolve the supercell artifacts from the physical features brought about
by randomness, one needs to ensure that the interaction between neighboring supercells,
which, to a good approximation, is proportional to the surface area of the supercell,
be small compared to the bulk properties of the imperfect PC, which can be taken 
to be proportional to the volume of the supercell. Hence, on purely dimensional grounds,
one can argue that the surface to volume ratio of the
supercell $1/L$, where $L$ is the linear size of the supercell, should be small compared to
the typical length scale of the problem, namely the wavelength of the bandgap.
We allowed the supercell size $N$ to vary from $N=2$ to 
$N\approx 9000$, and one can clearly see the supercell artifacts gradually diminishing 
while the bulk features become more prominent in the limit as $N\rightarrow \infty$.
On the other hand, 1D structures can have features,
such as a bandgap for any geometry and any refractive index contrast,
that are certainly not shared by 2D or 3D PCs. 

\begin{figure}[htbp]
\centering
\includegraphics[width=6cm]{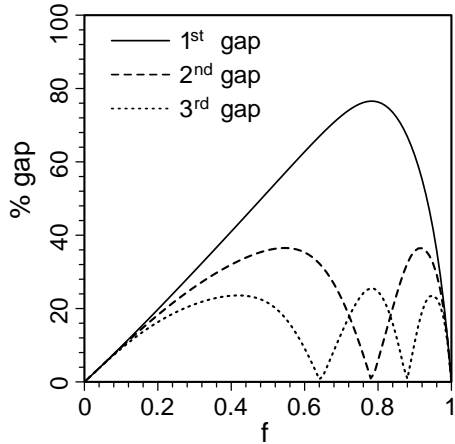}
\caption{\label{gapvsf}The relative gap width vs the filling ratio $f$ for a 1D PC made of 
slabs of alternating dielectric constant of $\epsilon_1=1$ and $\epsilon_2=13$. 
The lowest gap has a maximum for 
$f=1-\sqrt{\epsilon_1/\epsilon_2}=0.72$, which is the quarter-wave stack value. For this value 
of $f$, the even numbered gaps, the 2nd, 4th, etc, which are n general nonzero for an 
arbitrary value of $f$ are all closed.}
\end{figure}

The precise distribution of randomness in the geometry of a PC would
surely depend on the details of the specific manufacturing process.
In the interest of simplicity, we chose the simplest distribution,
the uniform distribution, in our study.
As the unit cell, we chose a
unit ``supercell" that consisted of up to $\sim 16000$ unit cells. The thicknesses
of the layers were
perturbed by a given percent, by adding random numbers chosen from a uniform
distribution. As the unperturbed structure, we chose the quarter-wave stack that
has, for a given dielectric contrast, the largest relative gap between 
the first and the second bands, as can be 
seen in Fig. \ref{gapvsf}. In what follows, we will consider this structure with a dielectric
contrast of 13 as our perfect PC.
For 1D PCs, one further has the luxury
of calculating the bandgaps using an analytical method\cite{yeh}. This approach 
also permits the calculation of the imaginary part of the wave vector 
in the forbidden gap region and allows a reliable assessment of the accuracy
of the plane wave method for the problem at hand.

We also investigated the transmission coefficient for a $250$ unit
cell quarter-wave stack structure. The transmission coefficient was
calculated by simply matching the boundary conditions for the
electric and the magnetic fields at each interface between the slabs 
in the multilayer structure.

\section{Density of states calculation with the PW method}
Maxwell's equations for waves propagating in the $x$ direction in a
medium with a dielectric constant $\epsilon(x)$ that depends only
on $x$, can be reduced to
\begin{equation}
\frac{\partial^2 E}{\partial x^2}-\frac{1}{c^2}\ \epsilon(x)\
\frac{\partial^2 E}{\partial t^2} = 0
\end{equation}
where $E$ is parallel to the slabs.
With $\epsilon(x)$ periodic along $x$ with lattice constant $a$, 
and translationally invariant along $y$ and $z$,
\begin{equation}
\nonumber
\epsilon(x) = \sum_g \epsilon(g) e^{igx}, 
\ \ \ \ \ \ \ {\rm with} \ \ \ \ \ \ \ 
\epsilon(g) = \frac{1}{a}\int_0^a \epsilon(x) e^{-igx}\, dx
\end{equation}
where $ g=m 2 \pi/a$, is a reciprocal lattice vector with  
$m=0, \mp1, \mp2, \ldots$, and $E(x)$ can be written as
\begin{equation}
E(x) = e^{ikx}\ \sum_g E(g) e^{igx}
\end{equation}
where $-\pi/a<k<\pi/a$. For a given $k$, this
yields an $\infty$-dimensional generalized eigenproblem
\begin{equation}
Q^2 E = \frac{\omega^2}{c^2} \epsilon E
\end{equation}
or by multiplying both sides from the left by $Q\epsilon^{-1}$, one
obtains the ordinary eigenproblem
\begin{equation}
(Q\epsilon^{-1}Q) (QE) = \frac{\omega^2}{c^2} (QE)
\end{equation}
where $Q\equiv (k+g)\delta_{gg'}$, $\epsilon_{gg'}\equiv \epsilon(g-g')$,
and $\epsilon^{-1}$ is the inverse of the matrix $\epsilon$.
For a given value of $k$, a truncation of this $\infty$-dimensional 
ordinary eigenvalue problem
yields, by retaining only the $g$-vectors with $|g|<g_{{\rm max}}$,
the band structure $\omega_j (k)$ and the modes $E_{jk} (g)$.
We choose a structure where the dielectric constant alternates
between two values $\epsilon_1$ and $\epsilon_2$ each with
thickness $d_1$ and $d_2$, respectively.

The choice of the lattice constant $a=d_1 +d_2$ is not unique. 
Although the choice $a=d_1 +d_2$ is the most obvious and the most convenient, 
the lattice constant
 can be chosen as any integer multiple of $d_1 +d_2$,  $A\equiv Na$.
With a choice for $A$ with $N >1$, and
following the same formalism one can write 
$$ \epsilon(x)=\sum_G \epsilon(G) e^{i Gx}
\ \ \ \ \ \ {\rm with,} \ \ \ \ \ \
\epsilon(G) = \frac{1}{A}\int_0^A \epsilon(x) e^{-iGx}\, dx$$
and 
$$ E(x) = e^{iKx}\sum_G E(x) e^{i Gx}$$
where $G=m(2\pi/A)$, with $m=0, \mp1, \mp2, \ldots$ and
$-\pi/A<K<\pi/A$. Clearly, to get results with the same level of accuracy
as before, {\em i.e.} with $N=1$, one would now need to include
$N$ times as many plane waves in the expansion, which simply
increases the computational burden, both in terms of storage and
computing time. 
\begin{figure*}[htbp]
\centering
\includegraphics[width=18cm]{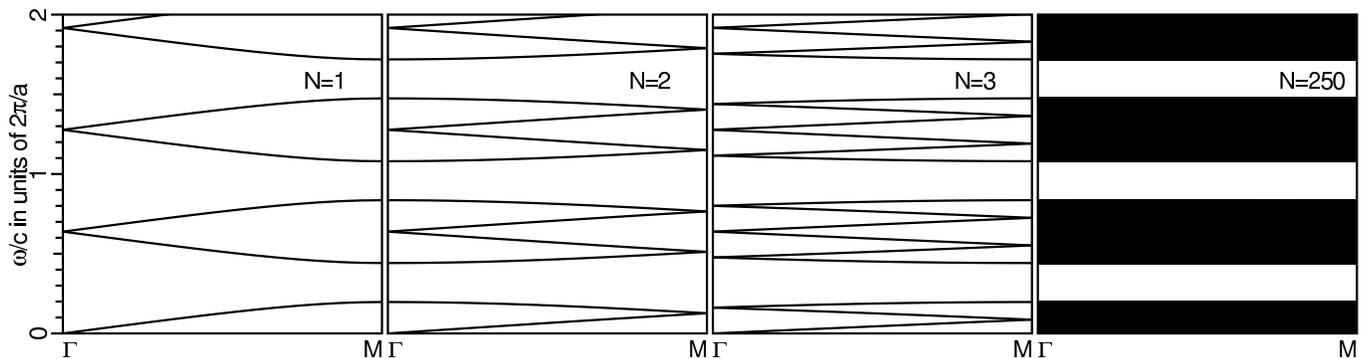}
\caption{\label{folding}The band structure of a perfect 1D PC with different choices
of supercell size $N$. The parameters of the structure are those of a quarter-wave stack, 
$\epsilon_1=13,\ \epsilon_2=1$, and $d_1/d_2=\sqrt{\epsilon_2/ \epsilon_1}$. 
The points $\Gamma$ and $M$ of the ``Brillouine zone" correspond to $K=0$ and $K=\pi/A$ 
respectively. When randomness is introduced, small gaps appear between each and every fold.}
\end{figure*}
The band structure for $N=1, 2, 3$ and $250$ are
displayed in Fig \ref{folding} for a perfect PC. i
The folding of the bands in the first
Brilloin zone for each $N$, makes the appearance of the bands
rather different for each case, although the DoS and the eigenfunctions
$E$ would be independent of the choice of
the supercell size. The frequency is plotted in units of
$2\pi/a$ for all cases, so the frequency scale is not affected with the
result that the bandgaps are at the same
frequency, as would be expected. To calculate the DoS, we 
choose a uniform mesh in $k$-space to calculate the bands and
then choose a small frequency window, $\Delta\omega$, and count
the number of modes whose frequencies fall within that window.

We add random perturbations to the thicknesses of the layers 
in the supercell such that
\begin{equation}
d_{1,2}= d_{1,2}^0\left[1+2p\left(u-\frac{1}{2}\right) \right]
\end{equation}
where $ d_{1,2}^0$ are the unperturbed values of the thicknesses of the layers,
{\em i.e.} the quarter-wave stack values,
$u$ is a uniformly distributed random number in the interval $(0,1)$.  
We control the amount of disorder by varying the percent randomness parameter $p$
between 0 and 1.  $p=0$ corresponds to perfectly
periodic structures, and $p=1$ corresponds to $100 \%$ fluctuation
where $d_1$, $d_2$ can range between $0$ and twice their
unperturbed values.
When disorder is introduced,
gaps appear between every fold for $N>1$.
\begin{figure}[htbp]
\centering
\includegraphics[width=7cm]{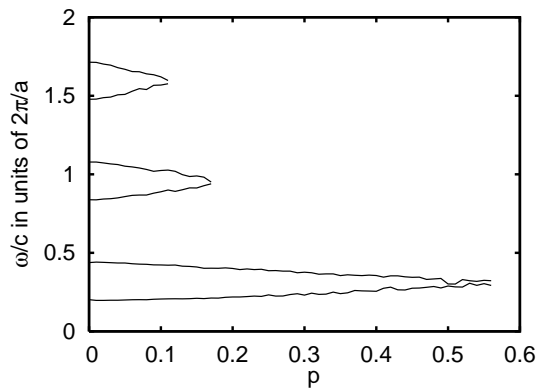}
\caption{\label{gapedgevsp}
The upper and lower band edges for the lowest three gaps
calculated with a supercell of size $N=1024$, as a function of the disorder parameter $p$.}
\end{figure}
In Fig. \ref{gapedgevsp} we plot the upper and lower limits for the lowest 
three bandgaps as a function
of $p$, the percent randomness with a supercell size of $N=1024$. 
Note that since for quarter-wave stack structures the even numbered gaps are
closed, the bandgaps in this figure are in fact the first, third and the fifth bandgaps
of a 1D PC with arbitrary values for the layer thicknesses.
The third gap centered at $\omega a/2\pi c=1.59$ closes 
around $p_3=0.1$, the second gap centered at $\omega a/2\pi c=0.96$ closes around 
$p_2=0.18$, and the lowest gap centered at $\omega a/2\pi c=0.32$ closes around 
$p_1=0.55$. The ratios of the critical values of randomness
$p_1:p_2:p_3$ agree well with the ratios of the corresponding center gap frequencies,
$\omega_3:\omega_2:\omega_1$. This can be understood using the simple argument that when
the random fluctuations in the thicknesses of the layers become comparable to the
wavelength of the gap center, the bandgap disappears since the destructive interference
responsible for the existence of the forbidden band depends on the long range 
periodicity at that scale. 
\subsection{Analytical Method}
\begin{figure*}[htbp]
\centering
\includegraphics[width=4.6cm]{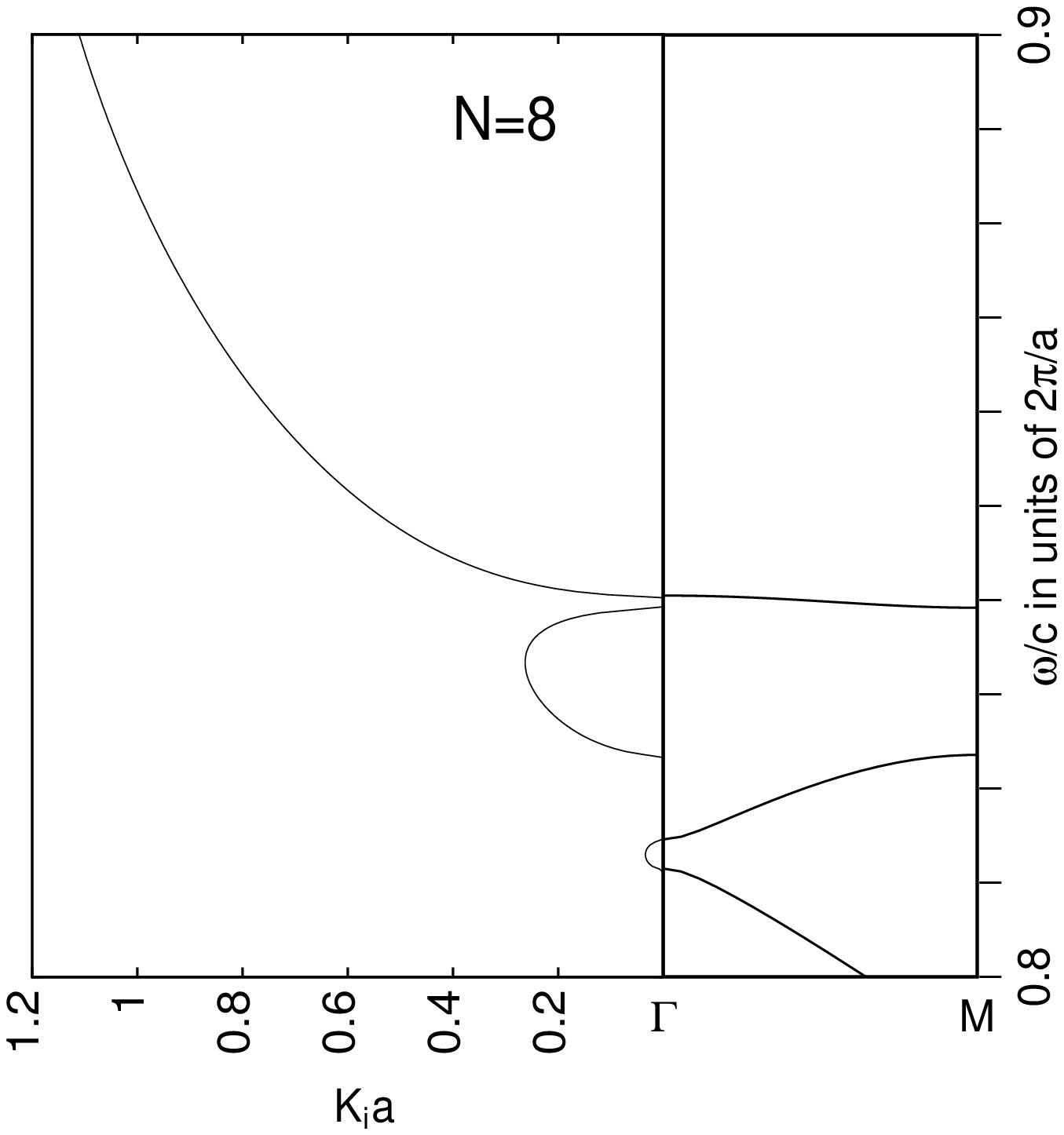}
\includegraphics[width=4.6cm]{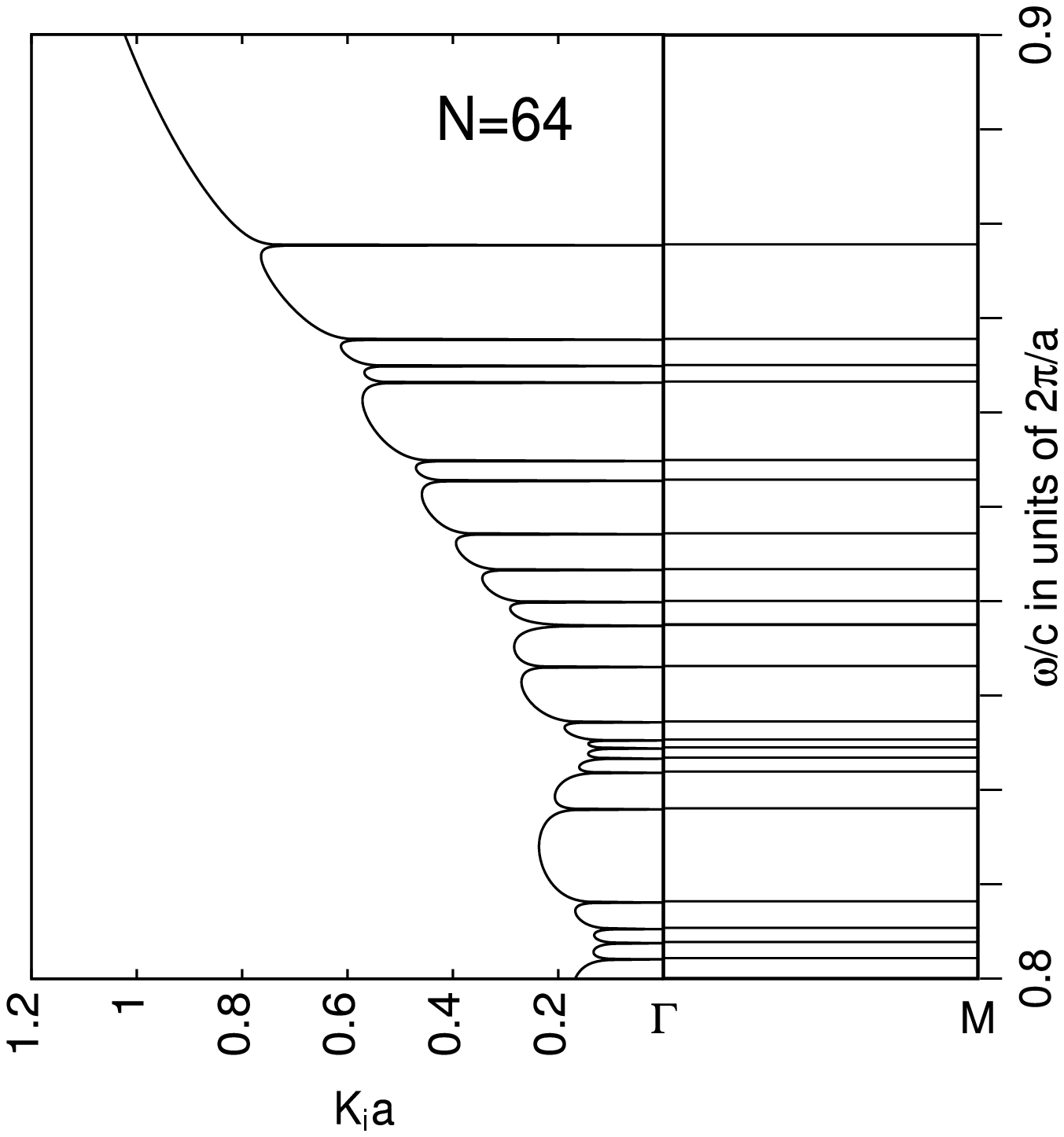}
\includegraphics[width=4.6cm]{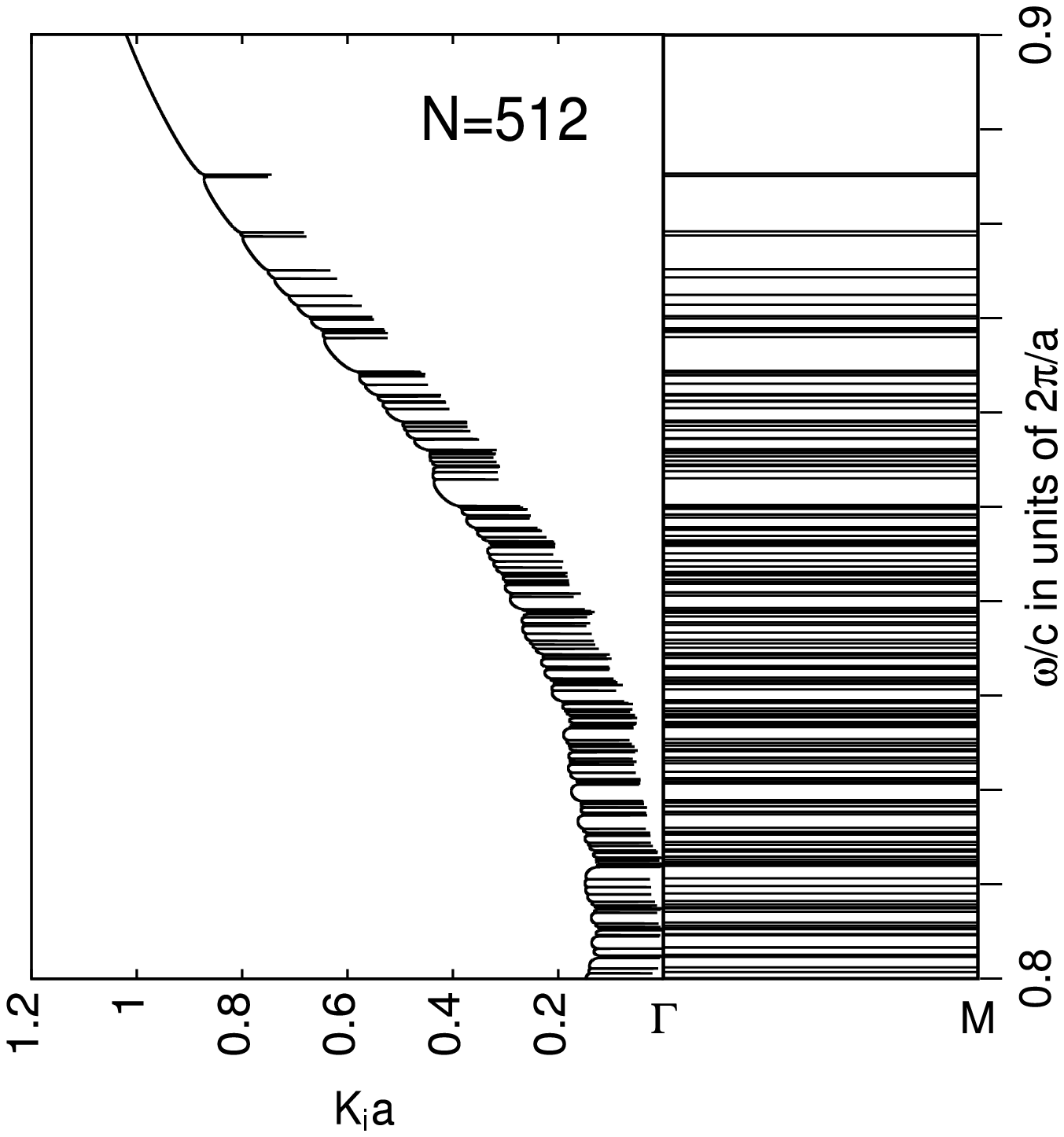}
\includegraphics[width=3.01cm]{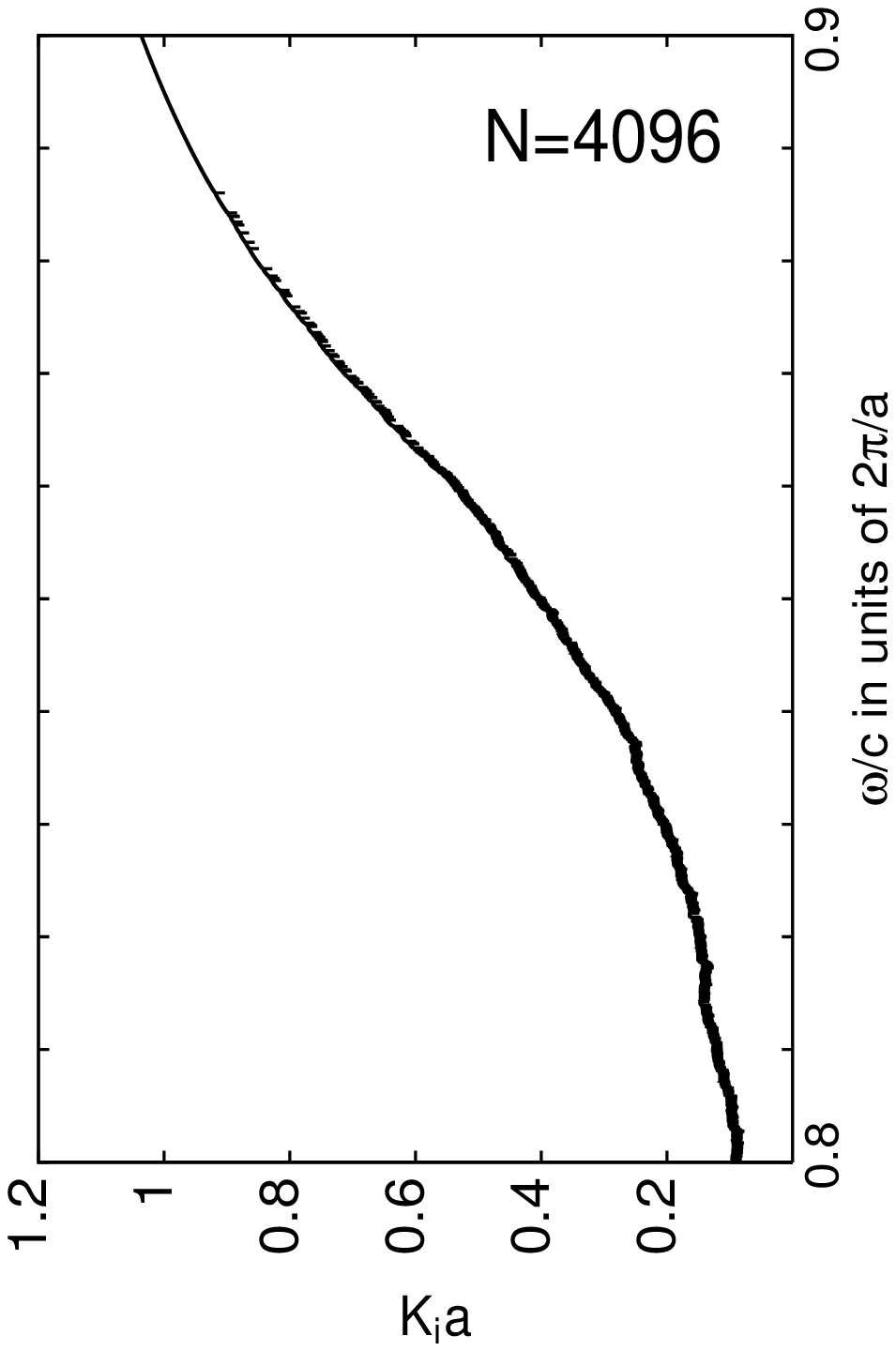}
\caption{\label{imbnd}The imaginary part of the wave vector $K_i$ and the band
structure for selected values of the supercell size $N$ for a randomness level of 10\%.
 The band structure
was calculated using the PW method and $K_i$ was calculated with the analytical
 method. Note the slight shift
of the bands due to PW convergence. As $N$ grows, the second bandgap that lies between
$0.83<\omega a/2\pi c<1.07 $ in the perfect PC is more and more populated with
transmission resonances, thereby narrowing the gap. For very large values of
$N$, the bandgap appears to settle down to $0.89\lesssim\omega a/2\pi c\lesssim 1.01 $. 
For $N=4096$, only $K_i$ is shown, as the PW method isn't practicali for such a 
large supercell. The cusps in the $K_i$ graph, which correspond to nearly flat bands,
are still identifiable. 
}
\end{figure*}
As discussed in detail in Ref\cite{yeh}, for $n$ dielectric layers
with thicknesses $d_1,\ldots,d_n$, with dieletric constants 
$\epsilon_1,\ldots,\epsilon_n$, for a given $\omega$, one can obtain 
the transfer matrix, defined by 
\begin{equation}
\label{Meq}
\left[\begin{array}{c}
E_0 \\
E_1
\end{array}\right]
=\left[\begin{array}{cc}
M_{11} & M_{12} \\
M_{21} & M_{22}
\end{array} \right]
\left[\begin{array}{c}
E_{2n} \\
E_{2n+1}
\end{array} \right]
\end{equation}
where 
\begin{equation}
M=D_0^{-1}\left(\prod_{l=1}^nD_lP_lD_{l}^{-1}\right)D_{n+1}
\end{equation}
with
\begin{equation}
\nonumber
D_l=\left[\begin{array}{cc} 
1 &1 \\
\sqrt{\epsilon_l} & -\sqrt{\epsilon_l}
\end{array} \right]
\ \ \ {\rm and}\ \ \ 
P_l=\left[\begin{array}{cc}
e^{i\sqrt{\epsilon_l}\omega d_l/c} &0\\
0&e^{-i\sqrt{\epsilon_l}\omega d_l/c}
\end{array} \right]
\end{equation}
Imposing the Bloch condition on the $E$-field, one obtains, 
\begin{equation}
\left[\begin{array}{c}
E_0 \\
E_1
\end{array}\right]
=e^{iKA}
\left[\begin{array}{c}
E_{2n} \\
E_{2n+1}
\end{array} \right].
\end{equation}
Comparing with Eq. \ref{Meq}, the eigenvalues of the transfer matrix are seen to be $e^{iKA}$. 
Then, for $t\equiv|(M_{11} +M_{22})/2|\leq 1$, $K$ is real and is given by
$K=(1/A)\cos^{-1}t$, while for $t>1$, $K$ is complex with $K_rA=\pi$ and 
$K_iA=-\ln(t-\sqrt{t^2 -1})$. $1/K_i$ is the decay length of the
evanescent mode, and is a measure of the strength of the bandgap. For 
finite PCs, it's desirable to have $K_iA\gg 1$ to have a 
significant drop in transmittance. The advantage of the exact method is that the
supercell size $N$ can be increased to values that are practically impossible
using the PW method.
While with the PW method, using 30 plane waves per unit cell of the supercell,
the memory requirements scale as $\sim (30N)^2$, and the time requirements scale
as $\sim (30N)^3$, the exact method requires a very small amount of memory.
\begin{figure}[htbp]
\centering
\includegraphics[width=8cm]{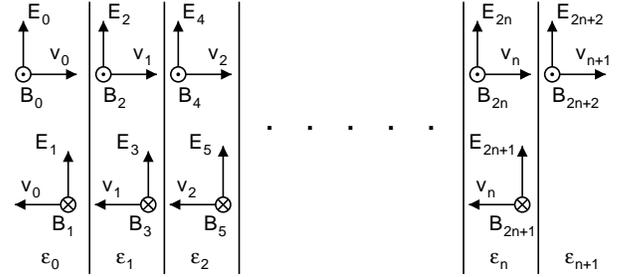}
\caption{\label{transfig}Fields used in transmittance and analytic badgap calculations.}
\end{figure}
The only disadvantage of the analytical method
over the PW method is that, while in the PW method one chooses a 
real $K$ and calculates the frequencies corresponding to that value of $K$,
in the analytical method, one chooses the frequency $\omega$ and calculates the
real and imaginary parts of $K$, $K_r$ and $K_i$, corresponding to that value of 
$\omega$. If the bands are nearly flat, as is the case for very large
supercell sizes, then one needs to sample the frequency interval of interest in
very tiny increments in $\omega$ in order to ``catch" a propagating mode.
Thus the computation time can become very large. Also for large values of $N$,
the transfer matrix $M$ can have very large elements so one requires very high
precision in order to calculate the transmission resonance frequencies. We used 
quadruple precision (128-bit) floating point variables and functions in the
Intel Fortran compiler in order to be able to 
resolve the transmission resonances for supercell sizes up to $N=8192$. 
For  large values of $N$, 
even 128-bit precision is not sufficient, with the result that $K_i$
cannot be made to completely vanish due to insufficient precision.
Nevertheless, the propagating modes appear as sharp cusps in the $K_i$ vs 
$\omega$ graph which can easily be identified (Fig. \ref{imbnd}). For $N\gtrsim 8000$, one
needs more than 128-bit precision to even see the cusps in (Fig. \ref{imbnd}).
For larger values of $N$, we used Mathematica for its arbitrary precision
capabilities. However, compared to compiled code, Mathematica is slower by several orders
of magnitude, so we had to stop at around $N=32000$.
\begin{figure}[htbp]
\centering
\includegraphics[width=5.2cm]{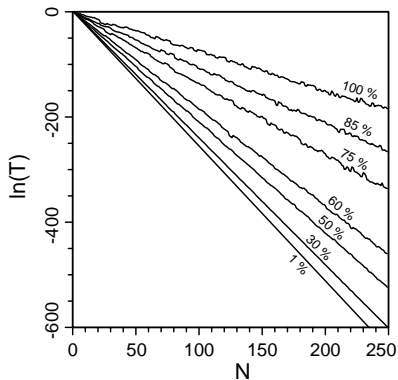}
\caption{\label{lntvsn}The dependence of $\ln T$ on $N$ for different values of
the randomness parameter $p$ for the center gap frequency of the first gap,
$wa/i2\pi c=0.32$. }
\end{figure}
To understand the {\em bulk} features of imperfect PCs using the supercell
method, one would need a large supercell, in fact the larger the better. As the
supercell size is increased, small bandgaps begin to appear over regions
that used to have propagating modes. One would normally expect the gaps
of the perfect crystal to gradually shrink in size, rather than have {\em more} gaps,
so this result seems somewhat puzzling at first sight. However, 
as the supercell size is increased, the statistical fluctuations decrease, and the pass bands
become increasingly more densely populated.
In Fig \ref{imbnd} we display the behavior of $K_i$ as $N$ is increased. 
As $N$ becomes larger 
what used to be a photonic bandgap becomes more and more populated with transmission
resonances, and the forbidden gap edges gradually approach each other, narrowing the
gap. It's possible that as $N\rightarrow\infty$,
the whole bandgap region will be populated, albeit extremely sparsely,  and instead of 
the bandgap, we will have
a region where the DoS is extremely small---but non-zero nevertheless. We were able to increase
up to $N=32768$ and the bandgap was reduced as $N$ became larger, although
the decrease for very large $N$ values was very small. To actually see the gap narrow even 
more would require an impractically large $N$.
\begin{figure}[htbp]
\centering
\includegraphics[width=4.18cm]{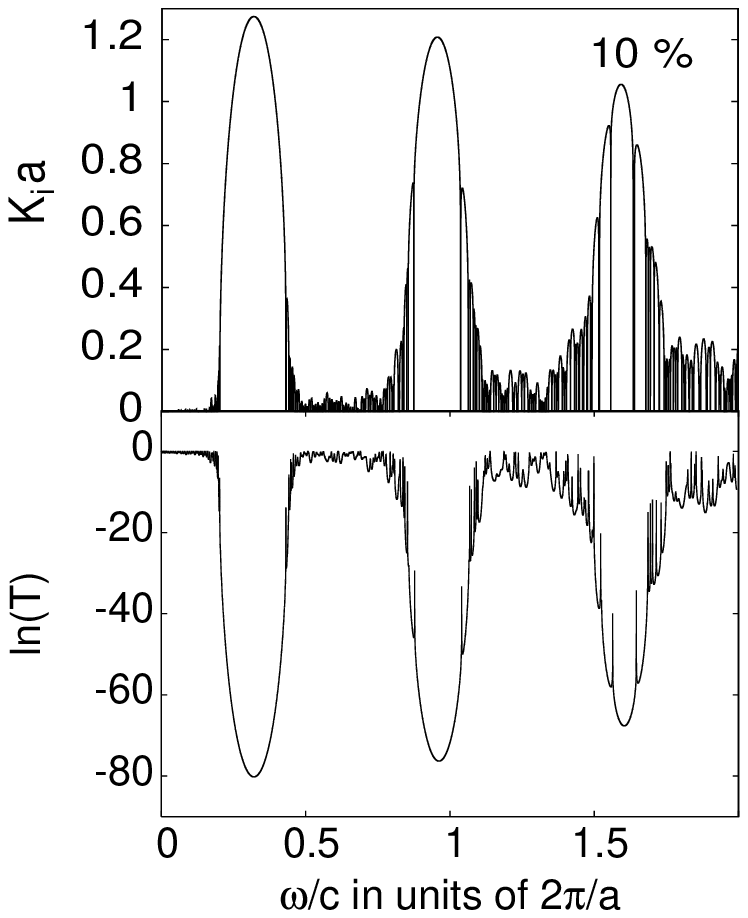}
\includegraphics[width=4.32cm]{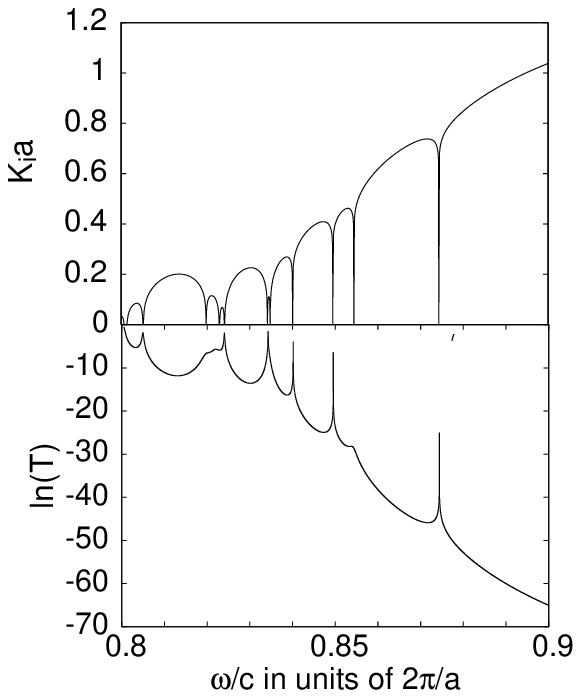}
\includegraphics[width=6cm]{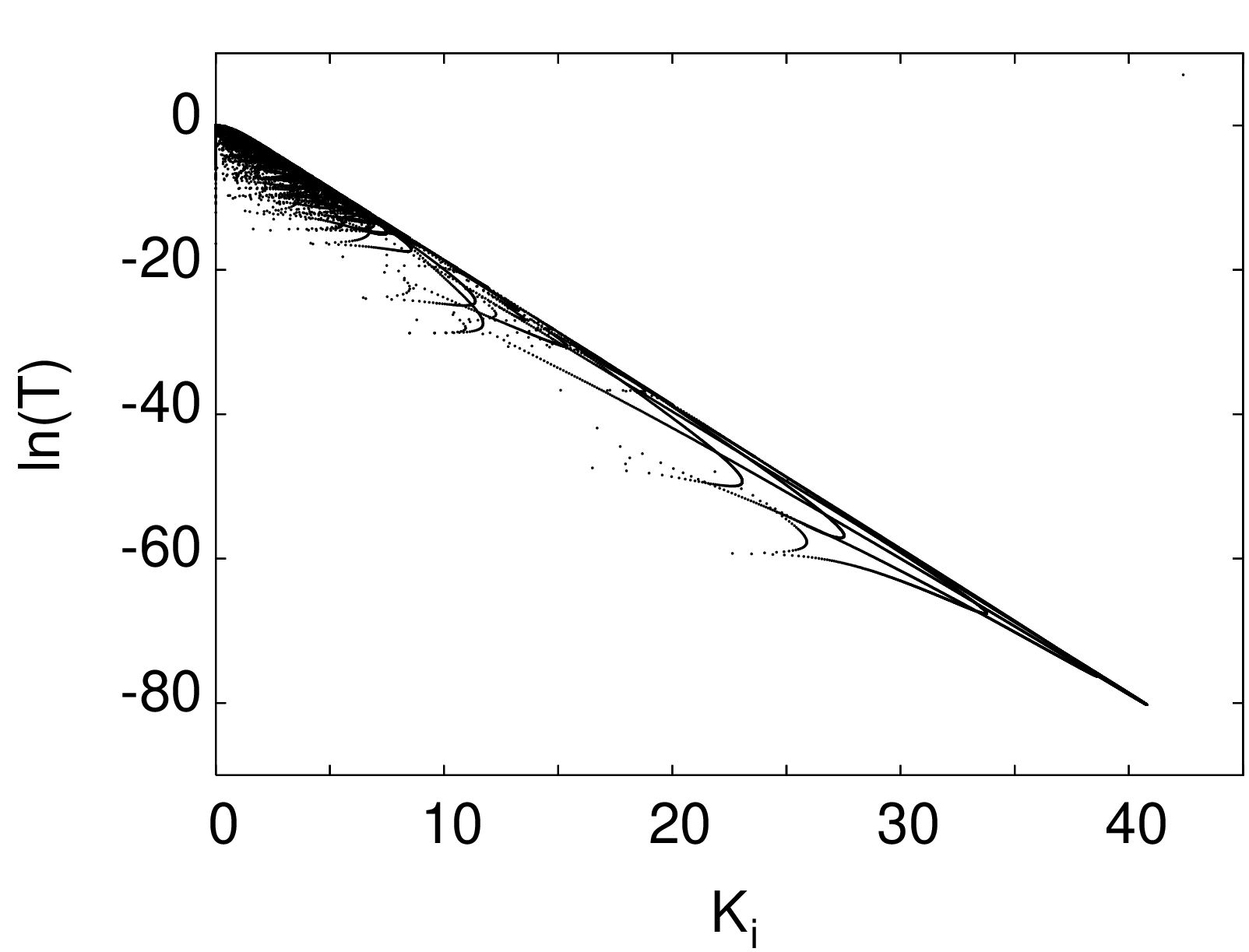}
\caption{\label{lntkivsw}(Upper left) $\ln T$ and $K_i$ vs $\omega$ for a supercell of size $N=32$
for a randomness level of 10\%. (Upper right) A closeup for $0.8<\omega a/2\pi c<0.9$ which 
contains the lower edge of the second gap $0.83<\omega a/2\pi c<1.08$ of the perfect PC.
(Lower) Scatterplot of $\ln T$ vs $K_i$ for the same structure.}
\end{figure}

The number of transmission resonances in any fixed frequency interval $\Delta \omega$
is proportional to $N$, so in the bulk limit with $N\rightarrow\infty$, any wave 
packet with a small, but nonzero frequency spread $\Delta \omega$ would contain
many transmission resonances and thus would be partly transmitted and partly reflected. 
Clearly in the bandgap regions of the perfect PC, the density of these transmission
resonances is extremely small, and these regions still appear to be bandgaps
with a large, but still finite, supercell.
Hence, it seems plausible to conclude that one cannot speak of a ``true bandgap" 
for imperfect PCs, but 
only of large depressions in the DoS, which, in practice, would serve the same purpose 
as {\em bona fide} bandgaps. For instance, for a cavity made of an ``impurity" embedded
in a PC, localized cavity modes would eventually leak out through
the PC ``walls" of finite thickness, regardless of how perfect the
PC walls are, bacause of the finite thickness of the walls. For such
an application, what is
important is that the lifetime of the cavity mode be much larger
than the relevant time scale. Since the lifetime of
the cavity mode is a function of the transmittance, for a given value of
transmittance, one would simply need to use thicker walls as the random
perturbations are increased.

\subsection{Transmittance}
\begin{figure*}[htbp]
\vspace{0cm}
\centering
\includegraphics[width=4.987cm]{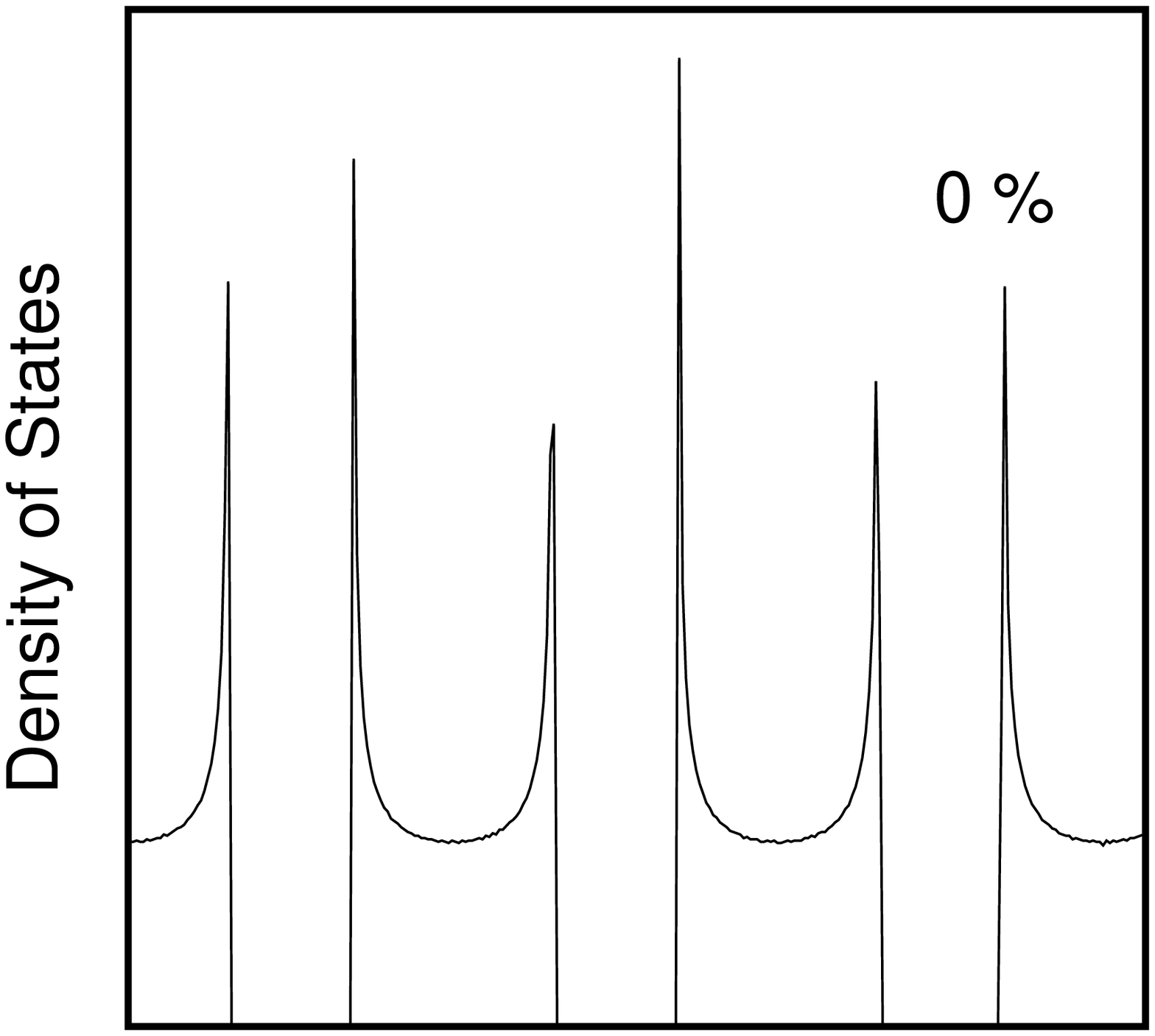}
\includegraphics[width=4.21cm]{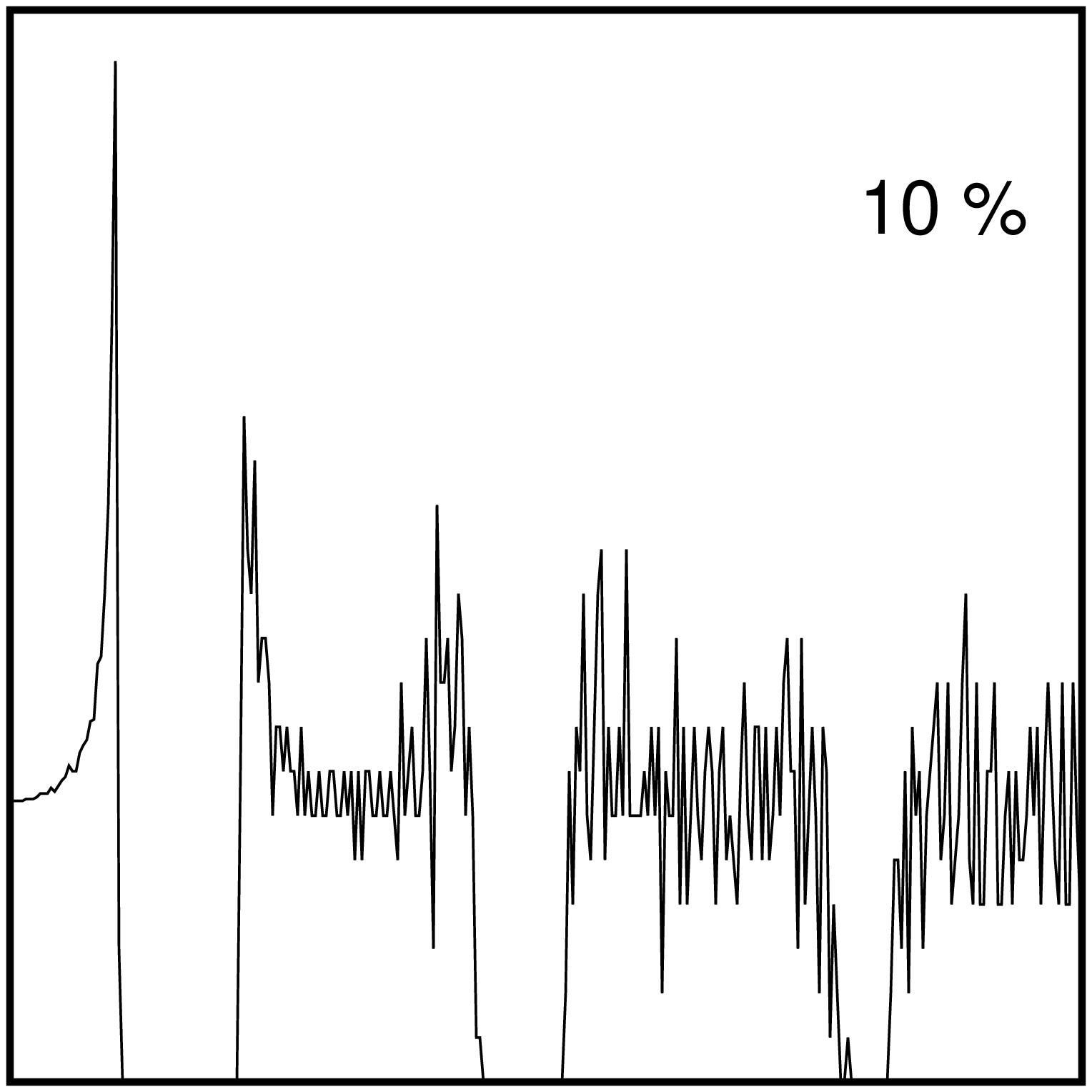}
\includegraphics[width=4.21cm]{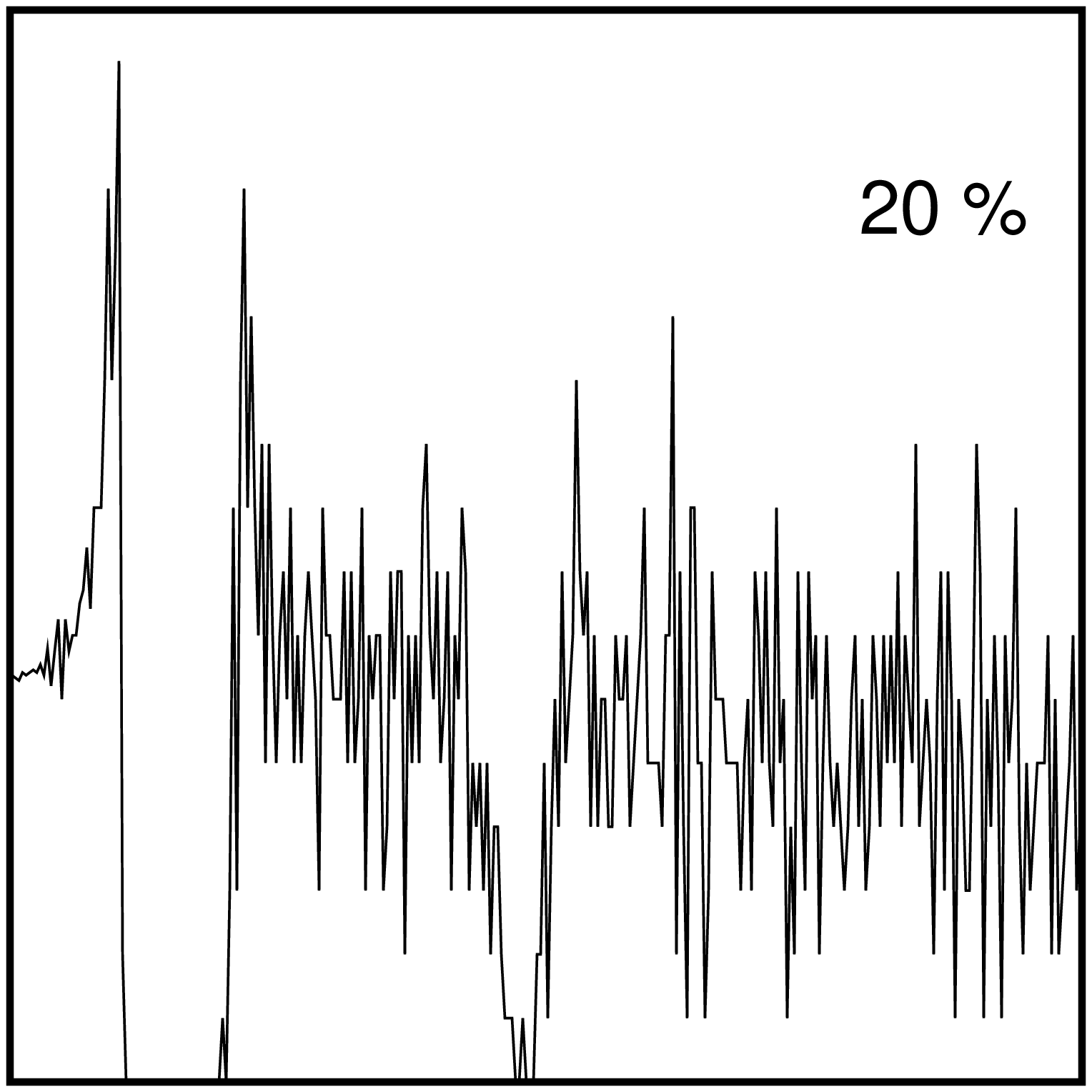}
\includegraphics[width=4.21cm]{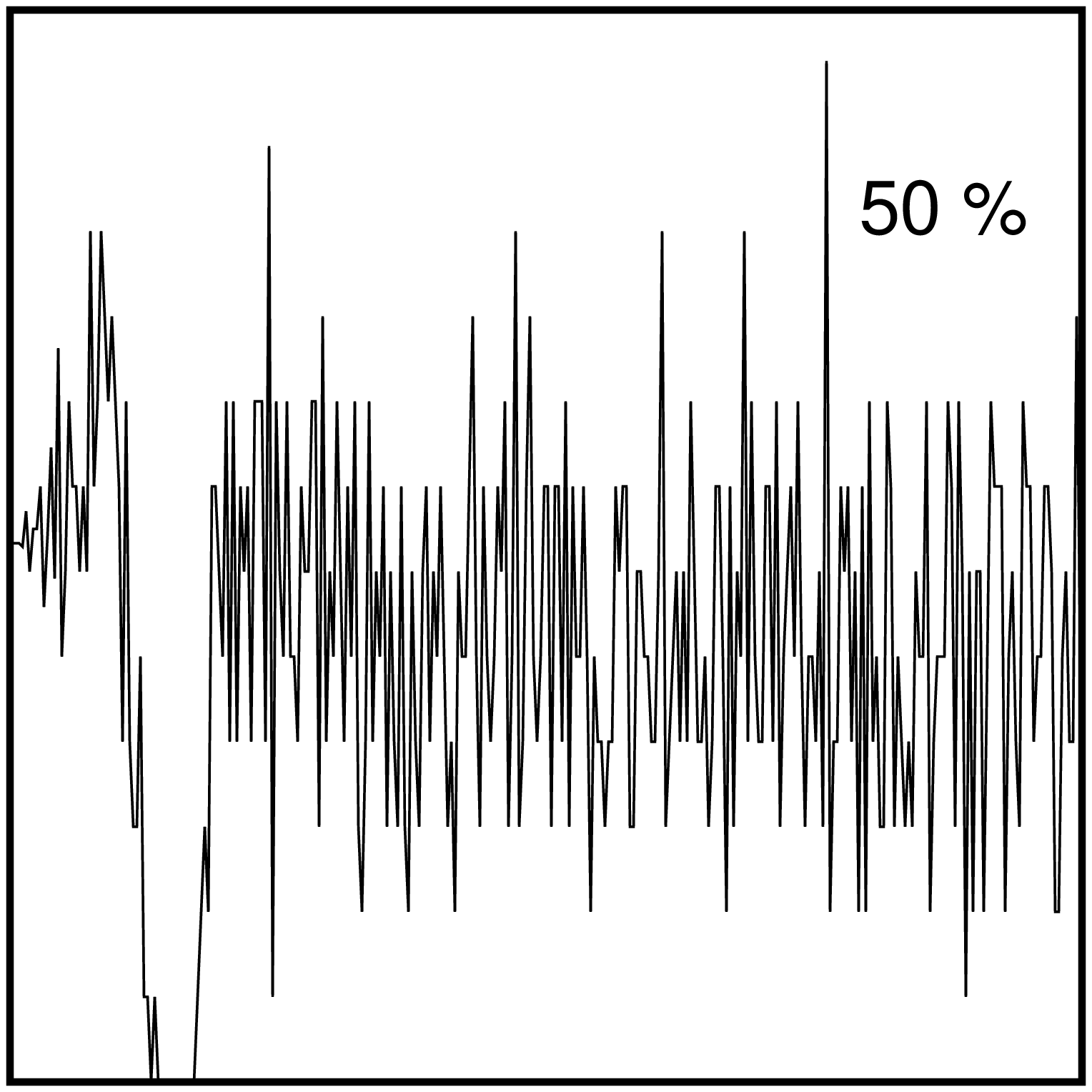}
\includegraphics[width=4.987cm]{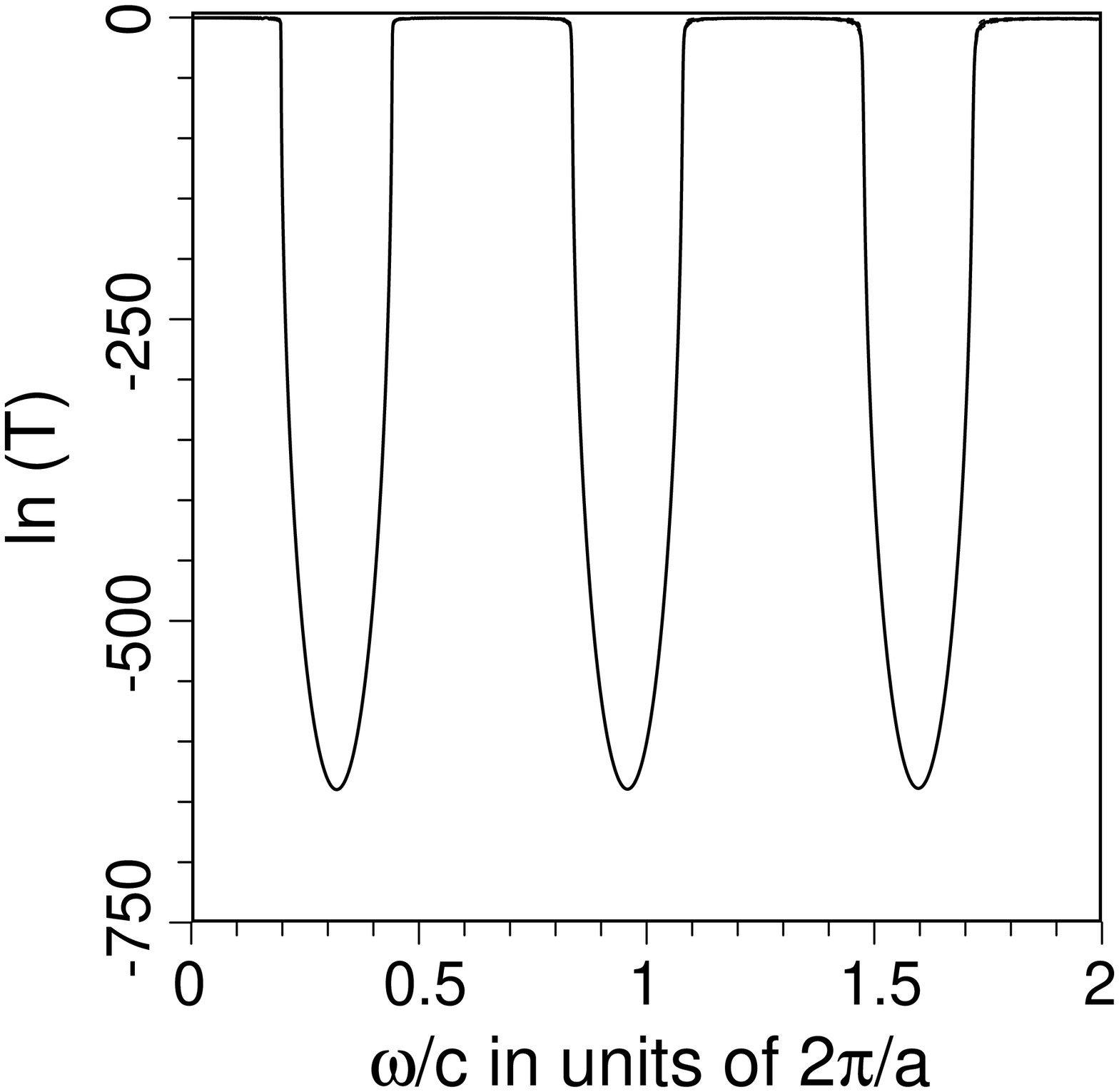}
\includegraphics[width=4.21cm]{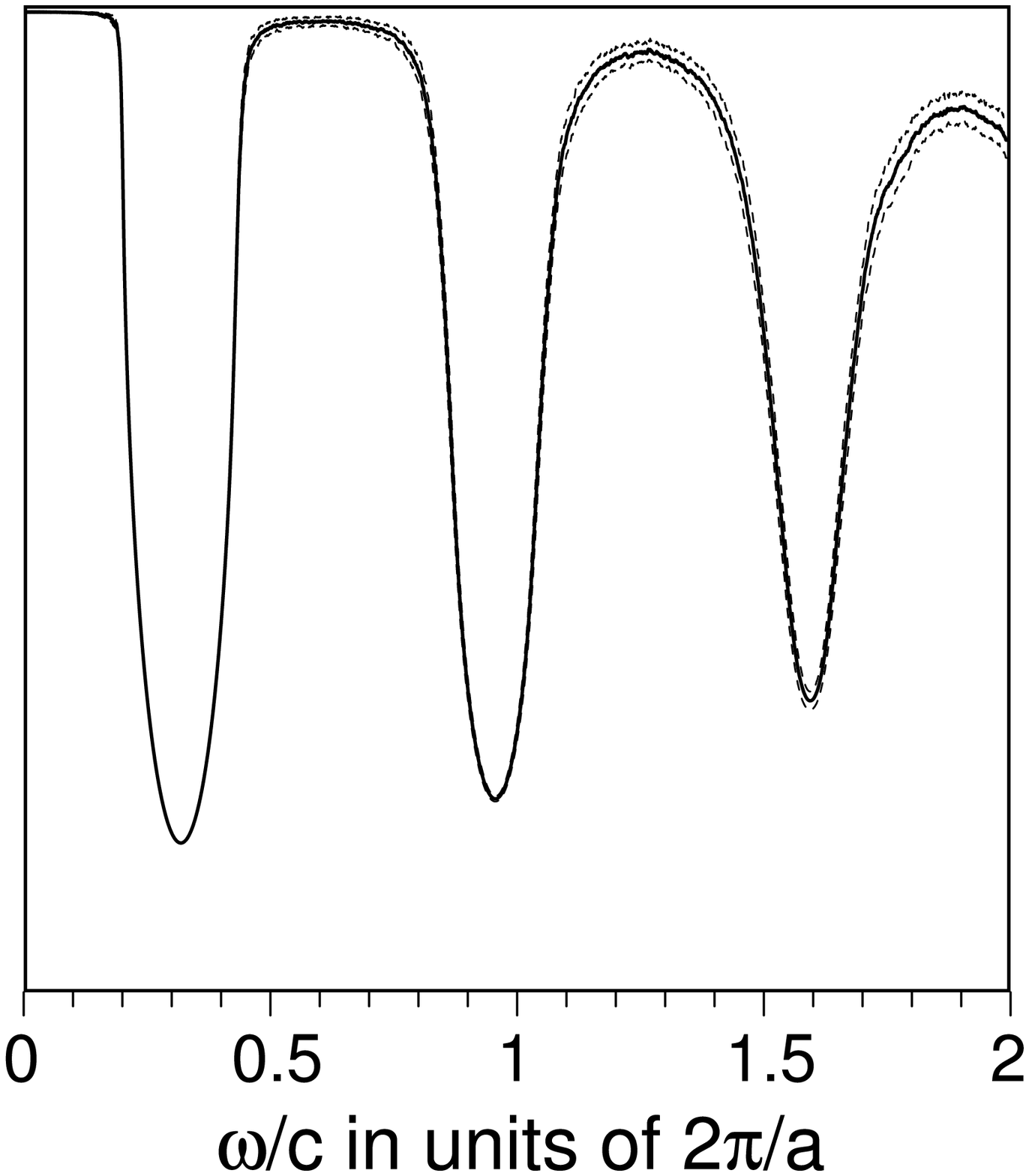}
\includegraphics[width=4.21cm]{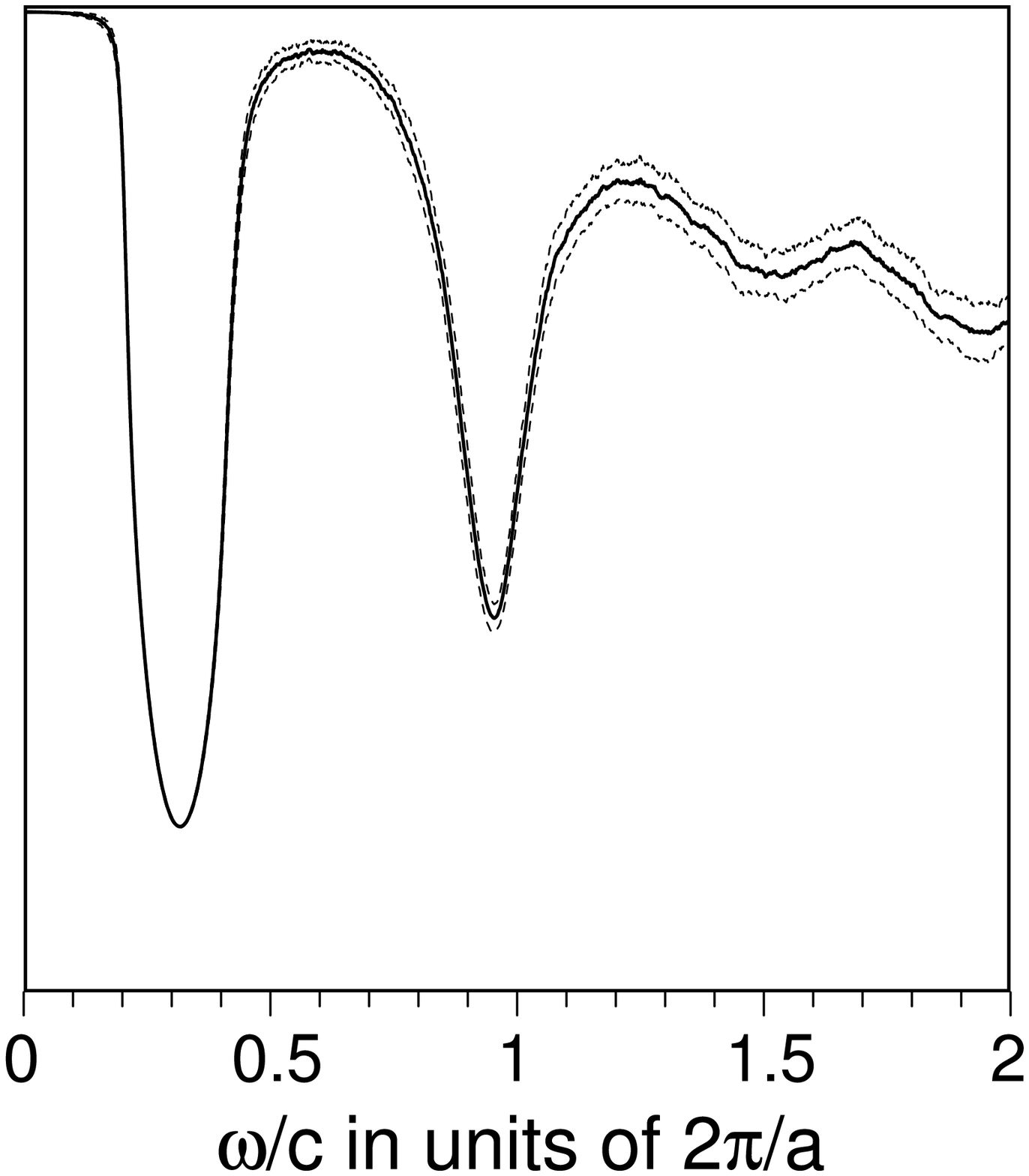}
\includegraphics[width=4.21cm]{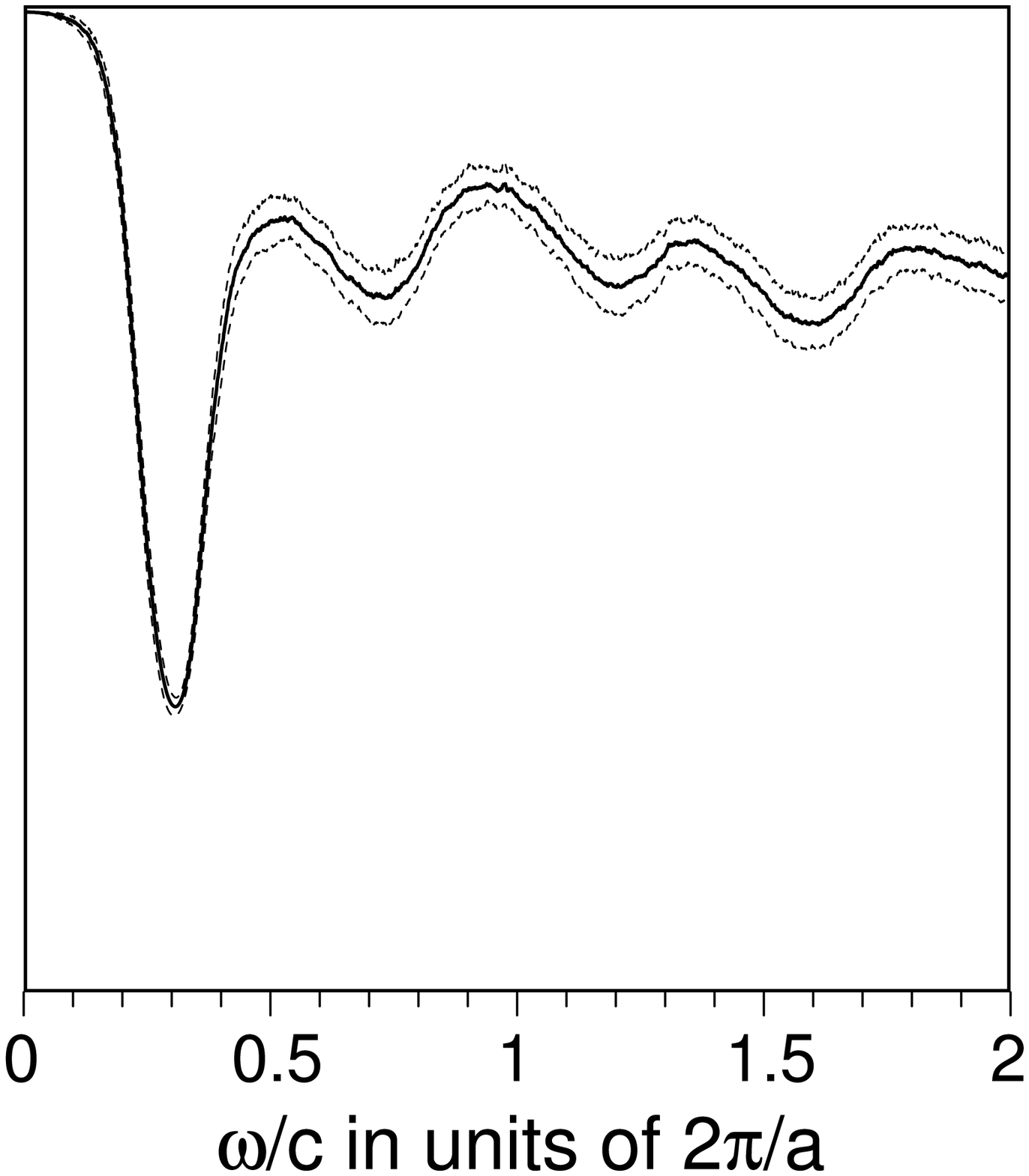}
\caption{\label{lntdosvsw}The DoS and $\ln T$ vs frequency for various levels of randomness.
 $\ln T$ (solid curve) is
the average for an ensemble of 100 random structures for each level of randomness.
Also shown as dashed curves are $\ln T\pm \sigma$, where $\sigma$
is the standard deviation of $\ln T$ for the ensemble used.}
\end{figure*}
Practical applications must necessarily use finite sized PCs, and for
such structures, a quantity of more relevance is the transmittance.
We calculate the transmittance by considering a PC of $N$ unit
cells, and each layer is perturbed as described earlier. The
transmission coefficient is calculated by imposing the boundary conditions
for $E$ $D$, $B$, and $H$ at each layer boundary (Figure \ref{transfig}). 
This yields a set of $2n+2$ linear equations for the unknowns 
$E_1, \ldots,E_{2n+2}$. Setting the incident field $E_0=1$, and 
assuming vacuum dielectric values for the incident and transmitted fields, 
$\epsilon_0=\epsilon_{n+1}=1$, one obtains, $R=E_1^2$ and $T=E_{2n+2}^2$ 
for the reflection and transmission coefficients, respectively. Alternatively,
one could also obtain $R$ and $T$ from the transfer matrix as detailed in Ref. \cite{yeh}
but with our approach, we can also obtain the $E_i$ within each layer.  
For a given frequency in the gap region, the dependence of $\ln T$ on the 
number of layers $N$ is approximately linear for 
all values of the randomness parameter $p$, as is the case \cite{yeh} for the perfectly 
periodic finite crystal (Fig. \ref{lntvsn}). However, depending on the level 
of randomness $\ln T$ can change by many orders of magnitude. In practice, 
this would mean that in
order to obtain a given value of transmittance using an imperfect PC, one would 
now have to use a thicker PC. 

Although there is a strong relationship between the transmittance of a finite imperfect
PC, and the modes of the superlattice formed by choosing the same exact finite PC as the
unit supercell, this relationship is not perfect in the sense that the existence 
of a propagating mode does not necessarily imply a large value for $T$. Conversely, 
the existence of a bandgap does not necessarily imply a low value for the 
transmittance $T$. A close examination of the $\ln T$ and $K_i$ vs $\omega$ plots in
Fig \ref{lntkivsw} will reveal that, on a large scale, the transmission has a dip 
where $K_i$ is large, and when  $K_i$ is small the transmittance is nearly unity. 
However, a closer look at a finer scale in Fig. \ref{lntkivsw}, one 
sees that  $\ln T$ can still be not as large as what one might expect from $K_i$.

Looking at
the plots of $\ln T$ vs $K_i$ Fig. \ref{lntkivsw}, one could argue that $K_i$ sets an upper limit
for $\ln T$. This upper limit seems to be a linearly decreasing function of $K_i$. That
the actual value of $\ln T$ can fluctuate below a particular value for a given $K_i$
can be understood if one considers the coupling of the incident plane wave to the
modes inside the PC. If, owing to the randomness, the propagating mode 
happens to have a small amplitude near $x_{n+1}$, it will not be transmitted effectively.
Similar coupling problems were reported by Robertson {\em et al} \cite{meade} for 2D PCs.
\section{Conclusion}
We studied the behavior of the photonic band gap and the transmittance for an imperfect PC
using the supercell method combined with both the plane wave method and the analytical 
method. We also studied the transmittance for an imperfect finite 1D photonic crystal
and have shown that the results we obtain in all cases are consistent. The bandgaps of the
perfect PC are replaced by a DoS that is extremely small to be detected even with extremely 
large supercells that consist of over 32000 unit cells. The higher frequency bandgaps
disappear first with the lowest gap closing at around a randomness level of $55\%$.
\section{Acknowledgements}
This work was supported by a grant from the Research Fund at Izmir Institute of Technology.
The band structure computations were performed on the 128-node Beowulf cluster at TUBITAK, 
the Scientific and Technological
Research Council of Turkey.

\end{document}